\documentclass[preprint,showpacs,showkeys]{revtex4}
\usepackage[T2A]{fontenc}
\usepackage[cp1251]{inputenc}
\usepackage[english]{babel}
\usepackage{amsmath}
\usepackage{amsfonts}
\usepackage{graphicx}
\usepackage{epsfig}
\usepackage{subfigure}
\usepackage{amssymb}
\usepackage[indent,small]{caption2}

\begin{document}

\title{Theory of Optical Bloch Oscillations in the Zigzag Waveguide Array}
\author{Babichenko V. S.$^1$, Gozman M. I.$^{1,2}$, Kagan Yu. M.$^1$, Polishchuk I. Ya.$^{1,2}$, Tsyvkunova E. A.$^3$}
\affiliation{$^1$ RRC Kurchatov Institute, Kurchatov Sq., 1, 123182, Moscow, Russia} %
\affiliation{$^2$ Moscow Institute of Physics and Technology, 141700, 9, Institutskii per., Dolgoprudny, Moscow Region, Russia} %
\affiliation{$^3$ Moscow Engineering Physics Institute, 115409, Kashirskoe highway, 31, Moscow, Russia} %

\begin{abstract}
The Bloch oscillations in the zigzag array of the optical
waveguides are considered. The multiple scattering formalism (MSF)
is used for the numerical simulation of the optical beam which
propagates within the array. The effect of the second-order
coupling which depends on the geometrical parameters of the array
is investigated. The results obtained within the MSF are compared
with the calculation based on the phenomenological coupling modes
model. The calculations are performed for the waveguides
fabricated in alkaline earth boro-aluminosilicate glass sample,
which are the most promising for the C-band %
($\lambda \sim 1530-1565\,\mathrm{nm}$).
\end{abstract}

\maketitle


\section{Introduction}

\label{Sec:Intro}


The rapid development of technologies for production of optical
waveguides provides a possibility of optical simulation of quantum
phenomena inherent in the condensed-matter physics
\cite{Review_Longhi2009}, such as
Bloch oscillations %
\cite{Pertsch1998_Theor,Pertsch1999,Morandotti1999,Pertsch2002,Chiodo2006, Gradons_Zheng2010}, %
Zenner tunneling %
\cite{Trompeter2006,Dreisow1_2009,Wang1_2010}, %
Anderson localization %
\cite{AndLoc_Schwartz2008,AndLoc_Lahini2008}, %
dynamic localization %
\cite{DinLoc_Garanovich2007,DinLoc_Dreisow2008}. %
Periodic arrays of optical waveguides are a special case of
photonic crystals. Such systems are of interest due to the band
structure of their optical spectrum that is similar to the band
structure in the electronic spectrum of ordinary crystals %
\cite{Joannopoulos, Busch}. %
Because of this, the behavior of light in photonic crystals is
analogous to the behavior of electron wave function in a periodic
potential of ordinary crystals. This fact is of considerable
interest for practical applications because it allows to steer the
optical signal effectively. Besides, it is interesting from a
purely scientific point of view, since the optical counterparts of
quantum phenomena can be observed directly.

\begin{figure}[tbp]
\centering
\includegraphics[width=0.4\textwidth]{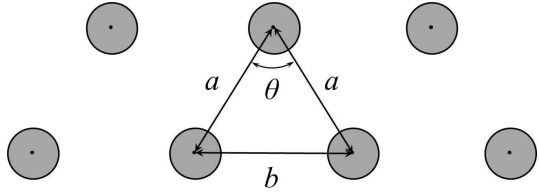}
\caption{Zigzag optical waveguide array.}
\label{Fig1_Scheme}
\end{figure}

Usually, the theoretical model of the light propagation in the optical
waveguides array takes into account only the next neighbors coupling (NNC).
This model is applicable for the particular case of the plane arrays, since
the high order coupling is too weak. But sometimes the second-order coupling
(SOC) should be taken into account. The most evident example of a system
with the significant SOC is the zigzag array (see Fig. \ref{Fig1_Scheme}).
For such a system the SOC depends on the angle $\theta$. Varying this angle,
we can make the SOC comparable or even greater then the NNC. The SOC can
influence the diffraction processes \cite{Zigzag_Dreisow2008} and forming of
the solitons in the case of the nonlinear optics \cite%
{ZigzagSolitons_Efremidis2002,ZigzagSolitons_Szameit2009}.


In this paper we investigate the influence of SOC on the optical
Bloch oscillations. The phenomenon of Bloch oscillations (BO)
takes place in the array with the monotonic change of the
refractive index of the waveguides as one passes from one
homogeneous waveguide to another. For the case of the plane array,
where only the NNC should be taken into account, the light beam
path possesses an oscillatory form. This effect is widely
investigated both experimentally and theoretically in
\cite{Pertsch1998_Theor,Pertsch1999,Morandotti1999,Pertsch2002,Chiodo2006}.
But for the zigzag array, where the SOC is considerable, the path
of the optical beam takes more complex form. This phenomenon,
known as the anharmonic Bloch oscillation (ABO), was
predicted in \cite{Wang2010} and experimentally confirmed in \cite%
{Dreisow2011}.


Numerical simulation of various phenomena in systems of interacting optical
waveguides, including BO and ABO, is usually based on a coupled modes model.
The parameters of this model are typically determined experimentally.

In our works \cite{We_OptEng2014,We_arXiv13105312} we proposed
another method of numerical simulation which requires no
additional data except of the geometrical and optical parameters
of the array (radii of the waveguides and distances between them,
the refractive indices of the waveguides and environment). Our
method uses the multiple scattering formalism based on the
macroscopic electrodynamics approach. In papers
\cite{We_OptEng2014,We_arXiv13105312} we presented the general
algorithm for calculating the path of the optical beam in an array
of waveguides.


In the present work we generalize the results obtained in
\cite{We_OptEng2014,We_arXiv13105312}. We apply the multiple
scattering formalism to the ABO calculation in the zigzag array.
The obtained picture of ABO is compared with the results of
calculation with use of the coupling mode model. For this purpose,
we derive the analytical formula for the path of the optical beam
in the zigzag array, using the coupling mode model with the SOC
taken into account. Besides, we use the MSF to obtain the
dependence of the coupling constant of the SOC as function of the
angle $\theta$ (see Fig. \ref{Fig1_Scheme}).


Recently a new method to fabricate low bend loss femtosecond-laser written
waveguides was developed [24, 25]. The waveguides are fabricated in alkaline
earth boro-aluminosilicate glass sample, so they are the most effective for
the C-band ($\lambda\sim 1530 - 1565\,\mathrm{nm}$). In this work we
consider a sample that can be fabricated by means of this technology.


This paper is organized as follows. In Sec. \ref{Sec:MSF} we describe the
multiple scattering formalism. In subsection \ref{Subsec:MSF01} we present
the algorithm for calculating the path of the optical beam in an array of
waveguides. In subsection \ref{Subsec:MSF02} we derive the equation of the
coupling modes model from the MSF and present the formulae for the coupling
constants. In Sec. \ref{Sec:Path} we represent the pictures of ABO obtained
by use of MSF for different $\theta$. We compare these results with the
paths of the optical beam described by the coupling modes model. Besides, in
Sec. \ref{Sec:Path} we present the dependence of the SOC coupling constant
on the angle $\theta$.

\section{Multiple scattering formalism}

\label{Sec:MSF}

Let us consider the array of $N$ parallel infinite cylindrical dielectric
waveguides. The axes of the waveguides are parallel to the $z$-axis. The
distance between the axes of the adjacent waveguides is denoted by $a$. All
the waveguides are assumed to possess the same radius $R$ but different
refractive indices $n_j$, $j$ being the waveguide number. The refractive
index of the environment is $n^0$. The permeability of the waveguide
material and the environment is unity.

Suppose that a guided mode with a frequency $\omega$ is excited within the
array. Then, all the components of the electromagnetic field are
proportional to $e^{-i\omega t+i\beta z}$.

Let us consider the field of the guided mode inside of the array. The field
of the guided mode is finite in any point into the waveguide. So, the
electromagnetic field inside of the $j$-th waveguide may be represented in
the form
\begin{equation}
\begin{array}{l}
\displaystyle \tilde{\mathbf{E}}_j(\mathbf{r})= e^{-i\omega t+i\beta
z}\,\sum\limits_{m=0,\pm1...} e^{im\phi_j}\, \Bigl(c_{jm}\,\tilde{\mathbf{N}}%
_{\omega_j\beta m}(\rho_j) -d_{jm}\,\tilde{\mathbf{M}}_{\omega_j\beta
m}(\rho_j)\Bigr), \medskip \\
\displaystyle \tilde{\mathbf{H}}_j(\mathbf{r})= e^{-i\omega t+i\beta
z}\,n_j\sum\limits_{m=0,\pm1...}e^{im\phi_j}\, \Bigl(c_{jm}\,\tilde{\mathbf{M%
}}_{\omega_j\beta m}(\rho_j) +d_{jm}\,\tilde{\mathbf{N}}_{\omega_j\beta
m}(\rho_j)\Bigr), \qquad\rho_j<R.%
\end{array}
\label{AppS_EHint}
\end{equation}
Here $\omega_j=n_j\omega$, and $\rho_j$, $\phi_j$ are the cylindrical
coordinates of the vector $\mathbf{r}-\mathbf{r}_j$, where $\mathbf{r}_j$ is
the coordinates of the axis of the $j$-th waveguide. The vector cylinder
harmonics $\tilde{\mathbf{M}}_{\omega_j\beta m}(\rho_j)$ and $\tilde{\mathbf{%
N}}_{\omega_j\beta m}(\rho_j)$ are defined as follows
\begin{equation}
\tilde{\mathbf{N}}_{\omega_j\beta m}(\rho_j) =\mathbf{e}_r\,\frac{i\beta}{%
\varkappa_j}\,J^{\prime}_m(\varkappa_j\rho_j) -\mathbf{e}_\phi\,\frac{m\beta%
}{\varkappa_j^2\rho_j}\,J_m(\varkappa_j\rho_j) +\mathbf{e}%
_z\,J_m(\varkappa_j\rho_j),   \label{tildeN}
\end{equation}
\begin{equation}
\tilde{\mathbf{M}}_{\omega_j\beta m}(\rho_j) =\mathbf{e}_r\,\frac{m\omega_j}{%
\varkappa_j^2\rho_j}\,J_m(\varkappa_j\rho_j) +\mathbf{e}_\phi\,\frac{%
i\omega_j}{\varkappa_j}\,J^{\prime}_m(\varkappa_j\rho_j),   \label{tildeM}
\end{equation}
where $\varkappa_j=\sqrt{\omega_j^2-\beta^2}$, $J_m(\varkappa_j\rho_j)$ is
the Bessel function, and the prime means the derivative with respect to the
argument $\varkappa_j\rho_j$.

Let us turn to the electromagnetic field outside of the array. This field is
the sum of the contributions of all the waveguides,
\begin{equation}
\mathbf{E}(\mathbf{r})=\sum\limits_{j=1}^N \mathbf{E}_j(\mathbf{r}), \qquad
\mathbf{H}(\mathbf{r})=\sum\limits_{j=1}^N \mathbf{H}_j(\mathbf{r}).
\label{Sum11}
\end{equation}
The contribution induced by the $j$-th waveguide and vanishing at $%
\rho_j\to\infty$ may be represented in the form
\begin{equation}
\begin{array}{l}
\displaystyle \mathbf{E}_j(\mathbf{r})= e^{-i\omega t+i\beta z}\sum\limits_m
e^{im\phi_j}\, \Bigl(a_{jm}\,\mathbf{N}_{\omega^{\prime}\beta m}(\rho_j)
-b_{jm}\,\mathbf{M}_{\omega^{\prime}\beta m}(\rho_j)\Bigr), \medskip \\
\displaystyle \mathbf{H}_j(\mathbf{r})= e^{-i\omega t+i\beta z} n^0
\sum\limits_m e^{im\phi_j}\, \Bigl(a_{jm}\,\mathbf{M}_{\omega^{\prime}\beta
m}(\rho_j) +b_{jm}\,\mathbf{N}_{\omega^{\prime}\beta m}(\rho_j)\Bigr),
\qquad\rho_j>R.%
\end{array}
\label{EX_J}
\end{equation}
Here $\omega^{\prime}=n^0\omega$. The vector cylinder harmonics $\tilde{%
\mathbf{M}}_{\omega_j\beta m}(\rho_j)$ and $\tilde{\mathbf{N}}%
_{\omega_j\beta m}(\rho_j)$ are
\begin{equation}
\mathbf{N}_{\omega'\beta m}(\rho_j)= \mathbf{e}_r\,\frac{i\beta}{%
\varkappa'}\,H'_m(\varkappa'\rho_j) -\mathbf{e}%
_\phi\,\frac{m\beta}{\varkappa'^2\rho_j}\,H_m(\varkappa'%
\rho_j)+\mathbf{e}_z\,H_m(\varkappa'\rho_j),   \label{N}
\end{equation}
\begin{equation}
\mathbf{M}_{\omega^{\prime}\beta m}(\rho_j)= \mathbf{e}_r\,\frac{%
m\omega^{\prime}}{\varkappa'^2\rho_j}\,H_m(\varkappa^{\prime}\rho_j)
+\mathbf{e}_\phi\,\frac{i\omega^{\prime}}{\varkappa^{\prime}}%
\,H^{\prime}_m(\varkappa^{\prime}\rho_j),   \label{M}
\end{equation}
where $\varkappa^{\prime}=\sqrt{\omega'^2-\beta^2}$, and $%
H_m(\varkappa^{\prime}\rho_j)$ is the Hankel function of the first kind.
Note that, for $\beta=0$, Eqs. (\ref{AppS_EHint}) and (\ref{EX_J}) transform
into the corresponding expressions in Ref. \cite{VanDeHulst}, however
different notations are used there.


Below we consider the simplest approximation to these equations, namely, the
\textit{zero-harmonic approximation}. This means that in Eqs. (\ref%
{AppS_EHint}) and (\ref{EX_J}) only the terms with $m=0$ are taken into
account. In this approximation there are two kinds of the guided modes,
namely the transverse magnetic (TM) and transverse electric (TE) modes. For
the TM-mode $b_{j0}=d_{j0}=0$, and for the TE-mode $a_{j0}=c_{j0}=0$. As an
example, let us consider the TM-modes.
\begin{equation}
\tilde{\mathbf{E}}_j(\mathbf{r})= c_j\,\tilde{\mathbf{N}}_{\omega_j\beta
0}(\rho_j), \qquad \tilde{\mathbf{H}}_j(\mathbf{r})= c_j\,n_j\,\tilde{%
\mathbf{M}}_{\omega_j\beta 0}(\rho_j) \qquad\rho_j<R.   \label{EHint_ZHA}
\end{equation}
\begin{equation}
\mathbf{E}_j(\mathbf{r})= a_j\,\mathbf{N}_{\omega^{\prime}\beta 0}(\rho_j),
\qquad \mathbf{H}_j(\mathbf{r})= a_j\,n^0\,\mathbf{M}_{\omega^{\prime}\beta
0}(\rho_j) \qquad\rho_j>R.   \label{EHex_ZHA}
\end{equation}
Here and below, $a_j$ and $c_j$ stand for $a_{j0}$ and $c_{j0}$, and the
factor $e^{-i\omega t+i\beta z}$ is omitted.

To derive the equations that determine the partial amplitudes $a_j$ and $c_j$%
, one should use the boundary conditions on the surface of every waveguide.
The boundary conditions on the surface of the $j$-th waveguide connect the
field $\tilde{\mathbf{E}}_j(\mathbf{r})$, $\tilde{\mathbf{H}}_j(\mathbf{r})$
inside the $j$-th waveguide and the field $\mathbf{E}(\mathbf{r})$, $\mathbf{%
H}(\mathbf{r})$ outside. In general, there are four independent boundary
conditions \cite{We_arXiv13105312}. However, for $m=0$ TM-modes and only two
boundary conditions are required:
\begin{equation}
\left[\mathbf{E}(\mathbf{R}_j)\right]_z= \left[\tilde{\mathbf{E}}_j(\mathbf{R%
}_j)\right]_z, \qquad \left[\mathbf{H}(\mathbf{R}_j)\right]_\phi= \left[%
\tilde{\mathbf{H}}_j(\mathbf{R}_j)\right]_\phi,   \label{MSF_BoundCond}
\end{equation}
here $\mathbf{R}_j$ is the radius-vector of a point on the surface of the $j$%
-th waveguide.

The boundary conditions on the surface of the $j$-th waveguide can be
expressed in the most convenient form if the contributions of all the
waveguides to the field outside the array are expressed in terms of the same
argument $\rho_j$. For this propose we apply the following relations:
\begin{equation}
\begin{array}{l}
\displaystyle \mathbf{N}_{\omega^{\prime}\beta 0}(\rho_l)\approx
U_{lj}(\omega,\beta)\, \tilde{\mathbf{N}}_{\omega^{\prime}\beta 0}(\rho_j),
\medskip \\
\displaystyle \mathbf{M}_{\omega^{\prime}\beta 0}(\rho_l)\approx
U_{lj}(\omega,\beta)\, \tilde{\mathbf{M}}_{\omega^{\prime}\beta 0}(\rho_j),
\qquad l\neq j,%
\end{array}
\label{MSF_Trans}
\end{equation}
where $U_{lj}(\omega,\beta)=H_0(\varkappa^{\prime}r_{lj})$, $H_0$ being the
Hankel function of the first kind and $r_{lj}$ being the distance between
the axes of the $j$-th and the $l$-th waveguides. The relations (\ref%
{MSF_Trans}) follow from the Graf theorem (see \cite{AbramowitzStegun}) in
the zero-harmonic approximation.

Thus, it follows from (\ref{EHex_ZHA}) that
\begin{equation}
\mathbf{E}_l(\mathbf{r})= a_j\,U_{lj}(\omega,\beta)\,\tilde{\mathbf{N}}%
_{\omega^{\prime}\beta 0}(\rho_j), \qquad \mathbf{H}_l(\mathbf{r})=
a_j\,n^0\,U_{lj}(\omega,\beta)\,\tilde{\mathbf{M}}_{\omega^{\prime}\beta
0}(\rho_j).   \label{EHex_ZHA01}
\end{equation}
Substituting (\ref{EHex_ZHA01}) to (\ref{Sum11}), one gets
\begin{equation}
\begin{array}{l}
\displaystyle \mathbf{E}(\mathbf{r})= a_j\,\mathbf{N}_{\omega^{\prime}\beta
0}(\rho_j) +\sum\limits_{l\neq j} a_l\,U_{lj}(\omega,\beta)\, \tilde{\mathbf{%
N}}_{\omega^{\prime}\beta 0}(\rho_j), \medskip \\
\displaystyle \mathbf{H}(\mathbf{r})= a_j\,n^0\,\mathbf{M}%
_{\omega^{\prime}\beta 0}(\rho_j) +\sum\limits_{l\neq j}
a_l\,n^0\,U_{lj}(\omega,\beta)\, \tilde{\mathbf{M}}_{\omega^{\prime}\beta
0}(\rho_j).%
\end{array}
\label{Sum11_01}
\end{equation}

So, the boundary conditions (\ref{MSF_BoundCond}) take the form
\begin{equation}
\begin{array}{l}
\displaystyle a_j\,H_0(\varkappa^{\prime}R) +\sum\limits_{l\neq j}
a_l\,U_{lj}(\omega,\beta)\,J_0(\varkappa^{\prime}R)= c_j\,J_0(\varkappa_j
R), \medskip \\
\displaystyle a_j\,i\frac{n^0\omega^{\prime}}{\varkappa^{\prime}}%
\,H_0^{\prime}(\varkappa^{\prime}R) +\sum\limits_{l\neq j}
a_l\,U_{lj}(\omega,\beta)\,i\frac{n^0\omega^{\prime}}{\varkappa^{\prime}}%
\,J_0^{\prime}(\varkappa^{\prime}R)= c_j\,i\frac{n_j\omega_j}{\varkappa_j}%
\,J_0^{\prime}(\varkappa_j R).%
\end{array}
\label{MSF_Syst01}
\end{equation}

Eqs. (\ref{MSF_Syst01}) lead to the following system of equations:
\begin{equation}
\frac{a_j}{\bar{a}_j(\omega,\beta)} -\sum\limits_{l\neq j}
U_{jl}(\omega,\beta)\,a_l=0,   \label{MSF_MainSyst}
\end{equation}
\begin{equation}
c_j=\bar{c}_j(\omega,\beta)\,a_j.   \label{MSF_cj0_aj0}
\end{equation}
Here
\begin{equation}
\bar{a}_j(\omega,\beta)= \frac{\varepsilon_j
\varkappa^{\prime}\,J_0^{\prime}(\varkappa_j R)\,J_0(\varkappa^{\prime}R)
-\varepsilon^0\varkappa_j\,J_0(\varkappa_j
R)\,J_0^{\prime}(\varkappa^{\prime}R)} {\varepsilon^0\varkappa_j\,J_0(%
\varkappa_j R)\,H_0^{\prime}(\varkappa^{\prime}R) -\varepsilon_j
\varkappa^{\prime}\,J_0^{\prime}(\varkappa_j R)\,H_0(\varkappa^{\prime}R)},
\label{MSF_a}
\end{equation}
\begin{equation}
\bar{c}_j(\omega,\beta)= \frac{\varepsilon^0\varkappa_j\{H_0(\varkappa^{%
\prime}R)\,J_0^{\prime}(\varkappa^{\prime}R)
-H_0^{\prime}(\varkappa^{\prime}R)\,J_0(\varkappa^{\prime}R)\}} {%
\varepsilon^0\varkappa_j\,J_0^{\prime}(\varkappa^{\prime}R)\,J_0(\varkappa_j
R)
-\varepsilon_j\varkappa^{\prime}\,J_0(\varkappa^{\prime}R)\,J_0^{\prime}(%
\varkappa_j R)}.   \label{MSF_c}
\end{equation}


\subsection{Method for the optical beam trajectory calculation.}

\label{Subsec:MSF01}

The system of equations (\ref{MSF_MainSyst}) describes the guided modes of
the array of waveguides. This system possesses the nontrivial solution only
if the determinant of the matrix of this system vanishes,
\begin{equation}
\det\left\|\frac{\delta_{jl}}{\bar{a}_j(\omega,\beta)} -U_{jl}(\omega,\beta)%
\right\|=0.   \label{MSF_det}
\end{equation}
This equation allows to obtain the propagation constants $\beta_n$ of the
guided modes for the given frequency $\omega$, $n$ being the number of a
guided mode. There are $N$ solutions of Eq. (\ref{MSF_det}).

Let $a_j(\beta_n)$ be the normalized solution of Eq. (\ref{MSF_MainSyst}), $%
\sum\limits_{j=1}^N\vert a_j(\beta_n)\vert^2=1$. The guided mode of the
frequency $\omega$ is a superposition of the modes with different $\beta_n$:
\begin{equation}
\begin{array}{l}
\displaystyle \mathbf{E}(t,\mathbf{r})=e^{-i\omega t}\, \sum\limits_n
C_n\,e^{i\beta_n z} \sum\limits_{j=1}^N a_j(\beta_n) \mathbf{N}%
_{\omega^{\prime}\beta_n 0}(\rho_j), \medskip \\
\displaystyle \mathbf{H}(t,\mathbf{r})=n^0\,e^{-i\omega t}\, \sum\limits_n
C_n\,e^{i\beta_n z} \sum_{j=1}^N a_j(\beta_n) \mathbf{M}_{\omega^{\prime}%
\beta_n 0}(\rho_j).%
\end{array}
\label{MSF_LinearSuperposition}
\end{equation}
The coefficients $C_n$ determine the superposition.

Let us introduce the modal amplitude
\begin{equation}
a_j(z) =\sum\limits_n C_n\,e^{i\beta_n z}\,a_j(\beta_n).   \label{iii}
\end{equation}
The functions $\mathbf{N}_{\omega^{\prime}\beta_n 0}(\rho_j)$, $\mathbf{M}%
_{\omega^{\prime}\beta_n 0}(\rho_j)$ vanish rapidly as $\rho_j$ increases.
So, the field near the $j$-th waveguide is mainly determined by the partial
amplitudes $a_j(\beta_n)$. So, the modal amplitude $a_j(z)$ represents the
behavior of the guided modes properly. The coefficients $C_n$ are obtained
from the boundary condition at $z=0$:
\begin{equation}
\sum\limits_n\,C_n\,a_j(\beta_n)=a_j(0).   \label{a4}
\end{equation}
The system of equations (\ref{a4}) allows to obtain the coefficients $C_n$
for given $a_j(0)$.

Below we suppose that the boundary conditions $a_j(0)$ possess the form of
the Gaussian beam,
\begin{equation}
a_j(0)=e^{-\frac{(j-j_0)^2}{\sigma^2}+ik_0 aj}.   \label{a5}
\end{equation}
This means that the external source approximately illuminates the ends of
the waveguides with the numbers $j_0-\sigma<j<j_0+\sigma$ and the phase
difference between the amplitudes taken at the ends of the nearest
waveguides is $k_0 a$.

Thus, the guided mode can be found as follows:

1) Calculate numerically the set of propagating constants $\beta_n$ using
Eq. (\ref{MSF_det});

2) Obtain the amplitudes $a_j(\beta_n)$ for every $\beta_n$ using Eq. (\ref%
{MSF_MainSyst});

3) For the given boundary conditions find the coefficients $C_n$ using Eq. (%
\ref{a4});

4) Calculate the function $a_j(z)$ by means of Eq. (\ref{iii}).


\subsection{Derivation of the coupled modes equation}

\label{Subsec:MSF02}


Suppose that the refractive index is $n_j=n_0+\delta n\times j$, $\delta
n\ll n_0$. Then, the solutions $\beta_j^{(0)}$ of the equation
\begin{equation}
\frac{1}{\bar{a}_j(\omega,\beta_j^{(0)})}=0   \label{MSF02_01}
\end{equation}
may be represented in the following form:
\begin{equation}
\beta_j^{(0)}=\beta_0^{(0)}+\alpha\,j, \qquad\alpha\ll\beta_0^{(0)}.
\label{assumption}
\end{equation}
For the case of the weak coupling, $\left\vert\left(\beta_n-\beta_j^{(0)}%
\right) /\beta_j^{(0)}\right\vert\ll 1$. Then, one has
\begin{equation}
\frac{1}{\bar{a}_j(\beta_n)}\approx \left.\frac{\partial}{\partial\beta}
\frac{1}{\bar{a}_j(\beta)}\right\vert_{\beta=\beta_j^{(0)}}
\times\left(\beta_n-\beta_j^{(0)}\right).   \label{9-2}
\end{equation}
Then, Eq. (\ref{MSF_MainSyst}) leads to the following result:
\begin{equation}
\Bigl(\beta_n-\beta_j^{(0)}\Bigr)\,a_j(\beta_k) -\sum\limits_{l\neq
j}\gamma_{lj}\,a_l(\beta_k)=0.   \label{MSF_CouplMode01}
\end{equation}
where
\begin{equation}
\gamma_{lj}(\beta_n)=\frac{U_{lj}(\omega,\beta_n)} {\displaystyle \left.%
\frac{\partial}{\partial\beta} \frac{1}{\bar{a}_j(\beta)}\right\vert_{\beta=%
\beta_j^{(0)}}}.   \label{9}
\end{equation}

For the zigzag array we can take into account the first-order and
second-order coupling only, i.e. $l=j\pm1$ and $l=j\pm2$. Due to the
symmetry, $\gamma_{j,j-1}=\gamma_{j,j+1}$ and $\gamma_{j,j-2}=\gamma_{j,j+2}$%
. Besides, since the waveguides differ insignificantly, one can neglect the
dependence of constants $\gamma_{j,j\pm1}$ and $\gamma_{j,j\pm2}$ on $j$.
Then, for the case of weak coupling and insignificant difference of the
waveguides one can neglect the dependence of $\gamma_{lj}(\beta_n)$ on $%
\beta_n$. Thus, Eq. (\ref{MSF_CouplMode01}) takes the form
\begin{equation}
\Bigl(\beta_n-\beta_j^{(0)}\Bigr)\,a_j(\beta_n) -\gamma_1\,\Bigl(%
a_{j-1}(\beta_n)+a_{j+1}(\beta_n)\Bigr) -\gamma_2\,\Bigl(a_{j-2}(%
\beta_n)+a_{j+2}(\beta_n)\Bigr)=0.   \label{MSF_CouplMode02}
\end{equation}
Here $\gamma_1$ and $\gamma_2$ stand for $\gamma_{j,j\pm1}$ and $%
\gamma_{j,j\pm2}$ correspondingly.

Now we can apply the obtained Eq. (\ref{MSF_CouplMode02}) to formulate the
equation for the function $a_j(z)$. It follows from (\ref{MSF_CouplMode02})
that the function $a_j(z)$ defined by Eq. (\ref{iii}) with any set of
coefficients $C_n$ satisfies the equation of the coupled modes.
\begin{equation}
\left(i\frac{d}{dz}+\beta_j^{(0)}\right)\,a_j(z) +\gamma_1\,\Bigl(%
a_{j-1}(z)+a_{j+1}(z)\Bigr) +\gamma_2\,\Bigl(a_{j-2}(z)+a_{j+2}(z)\Bigr)=0.
\label{MSF_CouplMode03}
\end{equation}


\section{Path of the Optical Beam}

\label{Sec:Path}

We apply the developed technique for calculating the optical beam in an
array represented in Fig. \ref{Fig1_Scheme}. The parameters taken for the
calculation correspond approximately to the parameters of the arrays of
waveguides reported in work \cite{Withford2013}.

The wavelength of the laser source in \cite{Withford2013} is $\lambda=1550~%
\mathrm{nm}$. We take the waveguide radius $R=5\lambda=7750~\mathrm{nm}$.
The refractive index of the environment is $n^0=1.4877$, and the refractive
index of the waveguide $j=0$ (the center of the array) is $n_0=n^0+5\times
10^{-3}$. These parameters also approximately corresponds to the experiments
reported in \cite{Withford2013}. For our calculations we assume that the
separation between the waveguides is $a=3R$, and a variation of refractive
indices between the nearest waveguides is $\delta n=n_j-n_{j-1}=5\times
10^{-6}$. We take the boundary conditions in the form (\ref{a5}) with $%
\sigma=4$ and $k_0=0$.

\begin{figure}[ptbh]
\centering
\includegraphics[width=0.6\textwidth]{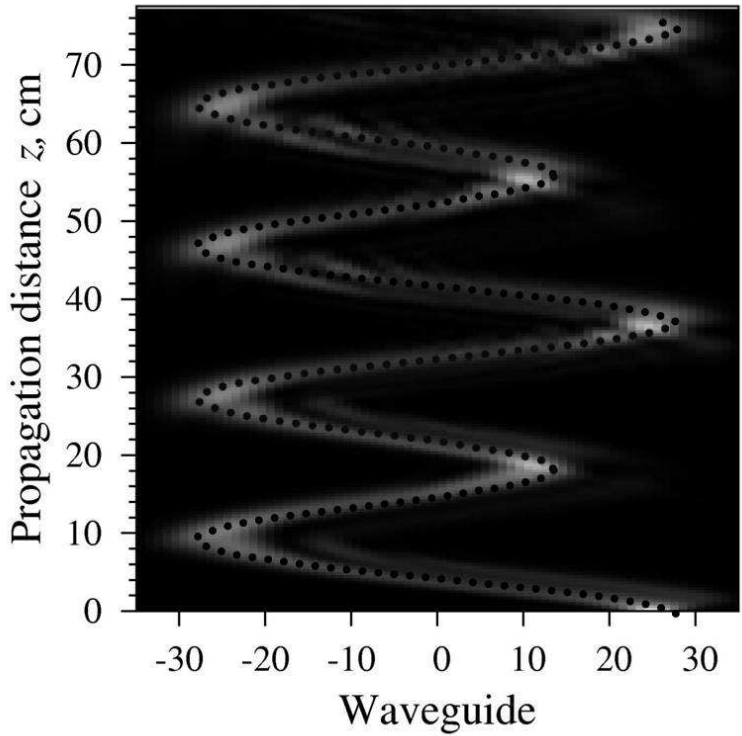}
\caption{$\protect\theta=50^\circ$}
\label{Fig50}
\end{figure}

\begin{figure}[ptbh]
\centering
\includegraphics[width=0.6\textwidth]{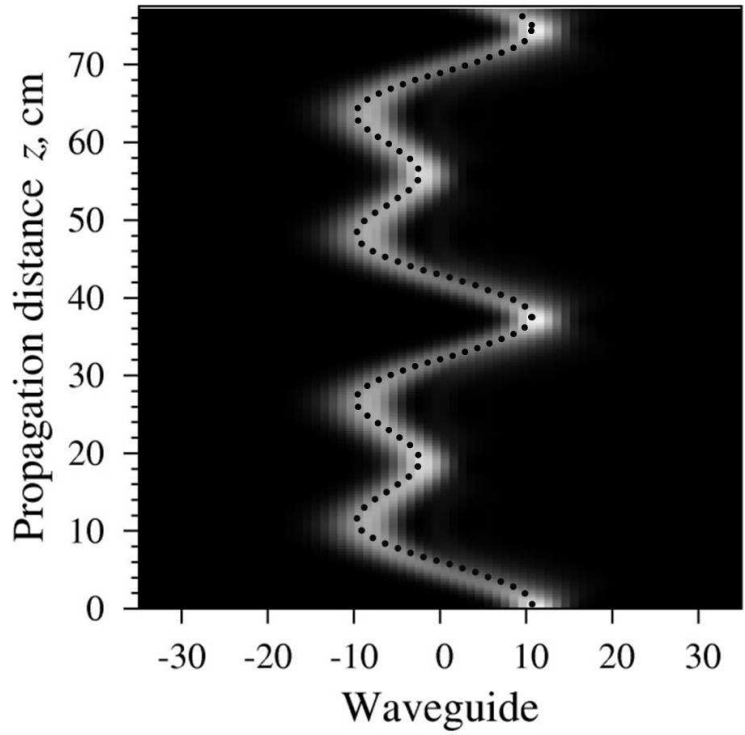}
\caption{$\protect\theta=60^\circ$}
\label{Fig60}
\end{figure}

\begin{figure}[ptbh]
\centering
\includegraphics[width=0.6\textwidth]{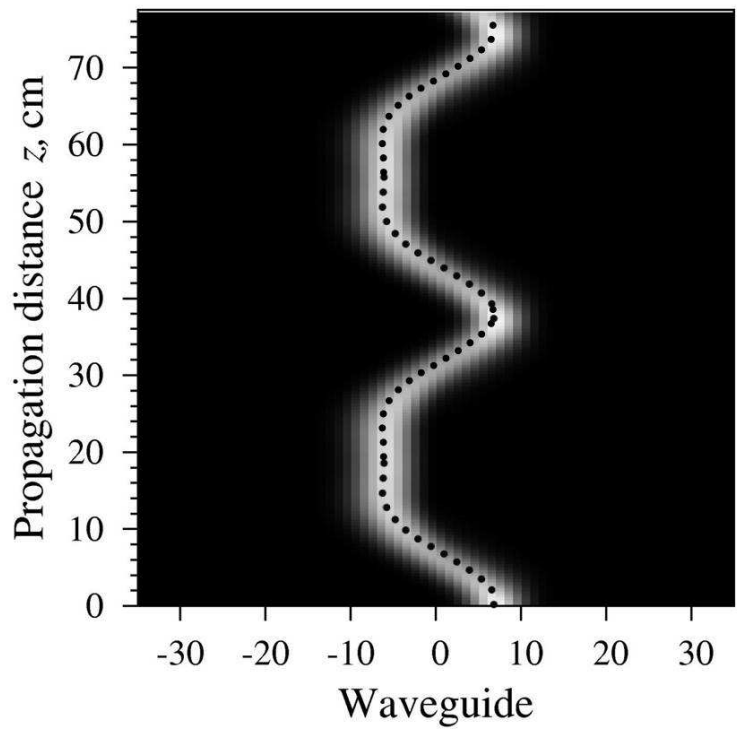}
\caption{$\protect\theta=70^\circ$}
\label{Fig70}
\end{figure}

\begin{figure}[ptbh]
\centering
\includegraphics[width=0.6\textwidth]{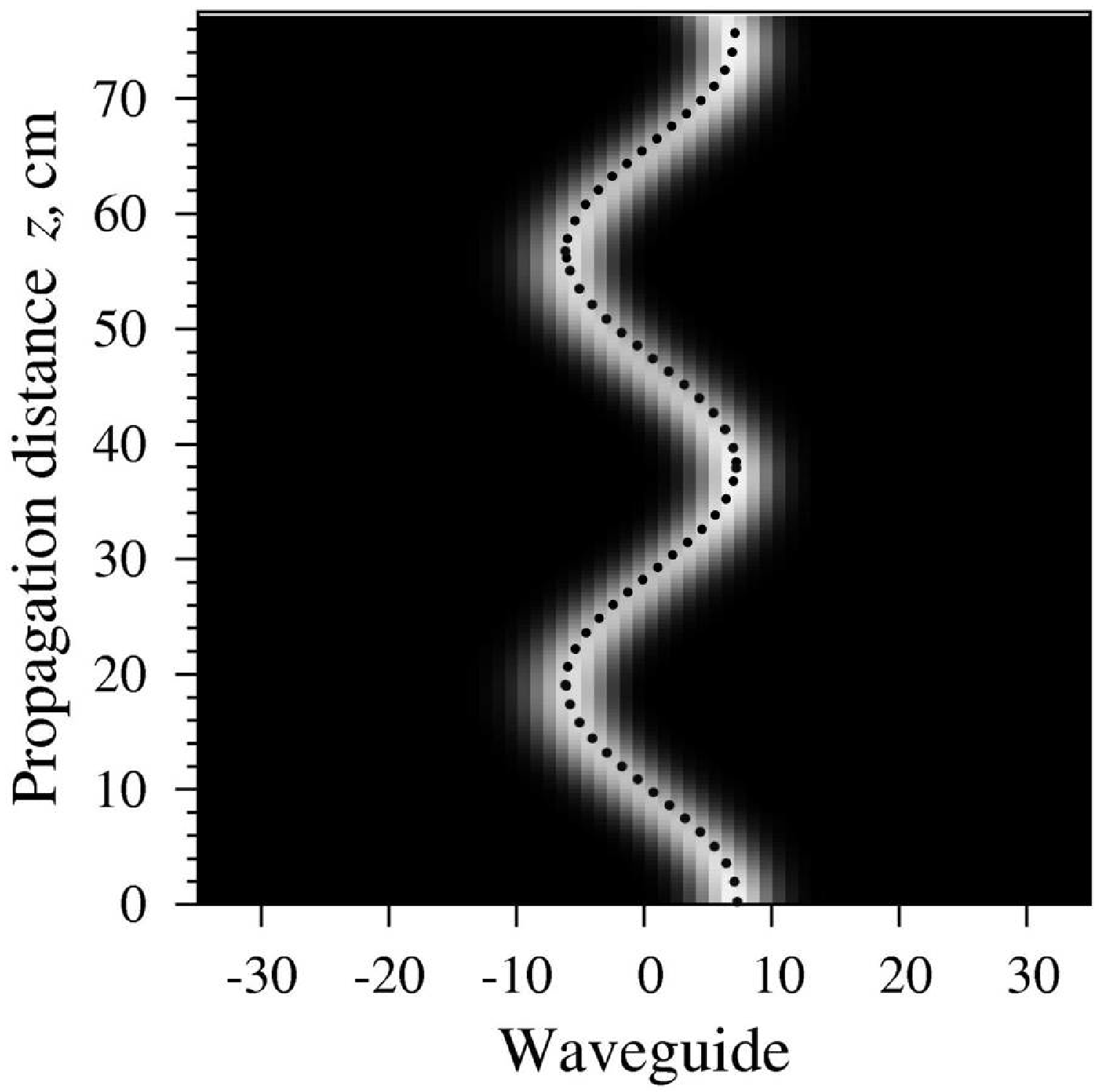}
\caption{$\protect\theta=180^\circ$}
\label{Fig180}
\end{figure}

We produce the calculation for different values of the angle $%
\theta=50^\circ,~60^\circ,~70^\circ,~180^\circ$ (the last case coincides
with the simple plane array). The results of the calculations are
represented in Figs. \ref{Fig50} -- \ref{Fig180}.

\begin{figure}[ptbh]
\centering
\includegraphics[width=0.6\textwidth]{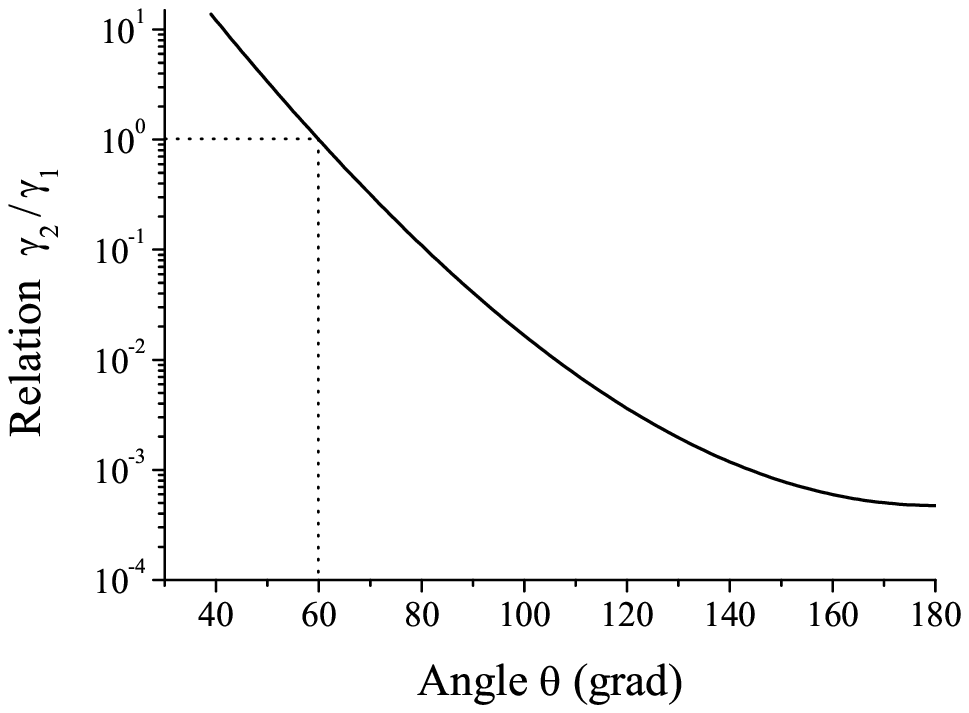}
\caption{The dependence of the relation $\protect\gamma_2/\protect\gamma_1$
on the angle $\protect\theta$.}
\label{Fig3}
\end{figure}

Every of these figures contains a dotted curve representing the trajectory
calculated by means of the coupled modes model. The parameter $\alpha$ is
found with the help of Eq. (\ref{MSF02_01}). The coupling constants $\gamma_1
$ and $\gamma_2$ for all the values $\theta$ are calculated by the use of
Eq. (\ref{9}). The values $\alpha$ and $\gamma_1$ do not depend on $\theta$
and they are the same for all the four Figs. \ref{Fig50} -- \ref{Fig180}: $%
\alpha=16.48~\mathrm{m}^{-1}$, $\gamma_1=-58.44~\mathrm{m}^{-1}$. But $%
\gamma_2$ depends on the distance $b$ between the second neighbors, so $%
\gamma_2$ decreases with increasing $\theta$: $\gamma_2(50^\circ)=-196.63~%
\mathrm{m}^{-1}$, $\gamma_2(60^\circ)=-58.44~\mathrm{m}^{-1}$, $%
\gamma_2(70^\circ)=-18.64~\mathrm{m}^{-1}$, $\gamma_2(180^\circ)=-0.0277~%
\mathrm{m}^{-1}$. The dependence of the relation $\gamma_2/\gamma_1$ on the
angle $\theta$ is depicted in Fig. \ref{Fig3}.

Note that for $\theta=60^\circ$ one has $\gamma_2=\gamma_1$. This is quite
expected result, since for this case $b=a$. Finely, note that for $%
\theta=180^\circ$ one has $\gamma_2/\gamma_1\sim 10^{-4}$. Thus, we confirm
that for the plane array of waveguides one can neglect the SOC.

\section{Conclusion}

\label{Sec:Conclusion}

In this paper we investigated the anharmonic Bloch oscillations in the
zigzag array of the optical waveguides. For this purpose we applied the
muliple scattering formalism based on the macroscopic electrodynamics
approach. For the simplicity, we used the zero-harmonic approximation taking
into account the TM-modes only. On the basis of the MSF, we developed the
numerical algorithm for calculating the optical beam path for the specified
boundary conditions at $z=0$. We chose the boundary conditions possessing
the form of the Gaussian wave packet. It was shown that for this case the
optical beam path takes the complicated periodic form.

This result is compared with the optical beam path predicted by means of the
coupled modes model with the second-order coupling taken into account. This
model contains some parameters, namely, namely, the difference between the
propagation constants of the adjacent waveguides $\alpha=\beta^{(0)}_j-%
\beta^{(0)}_{j-1}$ and the coupling constants $\gamma_1$ and $\gamma_2$ for
the first-order and the second-order coupling. These parameters were
calculated by means of MSF. We showed that the results of calculations by
two different methods are exactly the same.

Besides, we investigated the dependence of the coupling constant of the
second-order coupling $\gamma_2$ on the angle $\theta$. It was shown that $%
\gamma_2$ decreases quickly as the angle $\theta$ increases. For $%
\theta=180^\circ$ the value $\gamma_2$ becomes negligible. This result
confirms that for the plane array it is possible to produce the calculations
taking into account the next neighbors coupling only.


\subsection*{Acknowledgments}

The study is supported by the Russian Fund for Basic Research (Grants
13-02-00472a and 14-29-08165) and by the Ministry of Education and Science
of Russian Federation, project 8364.


\end{document}